\documentclass[%
 reprint,
nofootinbib,
 amsmath,amssymb,
 aps,
]{revtex4-2}
\usepackage{xcolor}
\usepackage{graphicx}
\usepackage{dcolumn}
\usepackage{bm}
\usepackage{hyperref}
\usepackage{mathtools}
\DeclareUnicodeCharacter{2212}{-}
\begin{document}

\preprint{APS/123-QED}

\title{Gravitational Slip parameter and Gravitational Waves in Einstein-Cartan theory}

\author{Maryam Ranjbar}
\email{maryamrnjbr96@gmail.com}
\affiliation{Department of Physics, Faculty of Sciences, Golestan University, Gorgan, Iran}

\author{Siamak Akhshabi}%
\email{s.akhshabi@gu.ac.ir}
\affiliation{Department of Physics, Faculty of Sciences, Golestan University, Gorgan, Iran}
\author{Mohsen Shadmehri}
\email{m.shadmehri@gu.ac.ir}
\affiliation{Department of Physics, Faculty of Sciences, Golestan University, Gorgan, Iran}

\date{\today}

\begin{abstract}
We study the evolution of scalar and tensor cosmological perturbations in the framework of the Einstein-Cartan theory of gravity. The value of the gravitational slip parameter which is defined as the ratio of the two scalar potentials in the Newtonian gauge, can be used to determine whether or not the gravity is modified. We calculate the value of slip parameter in the Einstein-Cartan cosmology and show that it falls within the observed range. We also discuss the evolution of the cosmic gravitational waves as another measure of the modification of gravity.

\end{abstract}

\maketitle

\section{\label{sec:intro}Introduction\protect}

 General Relativity (GR) has successfully predicted numerous phenomena, as corroborated by various studies \cite{intro1, intro2, intro3, intro4, intro5, intro6, intro7,intro17, introGR}. However, several mysteries persist in the realm of physics. The nature of the cosmological constant \cite{intro9, intro10, intro18, intro19, intro20, intro21}, dark energy \cite{intro9, intro10, intro8, intro11, intro16}, and dark matter \cite{intro12, intro13, intro14, intro15} are among the phenomena yet to be comprehensively understood. Various models have been proposed to address these unaccounted phenomena \cite{re1, re2, re3, re4, re5, re6, intro22, intro23, intro24, intro25, intro26,reDEHARO20231}.

One potential reason for these uncertainties is that GR, in the context of quantum mechanics, is fundamentally a classical theory. It is primarily tailored for the analysis of macroscopic phenomena. To address this limitation and extend the scope of GR to microphysical processes, various modified gravity theories have been proposed. Notably among them are those that factor in the torsion of spacetime, which is described as the non-symmetric part of the affine connection.

The simplest of such theories, Einstein-Cartan theory (ECT) \cite{intro27, intro28, intro29, intro44} has the same Lagrangian as in GR but the geometry of space-time is no longer Riemannian, but is a more general Riemann-Cartan space-time \cite{intro27, intro28, intro29, hehl-1974, hehl-1976}. In Einstein-Cartan theory, torsion is coupled to the spin angular momentum of the matter fields which acts as a source for gravitational interactions. Einstein-Cartan theory is well studied in the literature and has been used to study cosmological constant's sign problem \cite{cosmologicalconstantsign}, Hubble tension problem \cite{cosmologicalconstant}, Big Bang singularity \cite{intro32,intro33,intro34,intro35,intro36}, the emergent universe scenario \cite{intro39}, inflationary scenario \cite{intro37,intro38,intro40, intro43}, and Dark matter \cite{darkmatterect}.\\

 The detection of gravitational waves by LIGO has provided a unique opportunity to gain valuable insights into both astrophysics and cosmology, particularly in relation to the early universe \cite{intro6}. Within the framework of ECT, the presence of torsion can introduce notable changes in the generation and propagation of gravitational waves \cite{regwec1, regwec2}. Research presented in \cite{gwect1} has demonstrated that neglecting the consideration of torsion may lead to systematic errors when studying the propagation of gravitational waves. Furthermore, as highlighted in \cite{gwect2}, ECT can yield significantly different descriptions of gravitational wave fluctuations compared to those derived from General Relativity (GR). Given that the effects of torsion and spin during the early stages of the universe are assumed to be significant \cite{torsionearlyuniverse}, investigating the behavior of GW through the ECT not only allows us to test the validity of this model, but also has the potential to reveal novel understandings regarding the nature of gravity.
 
Typically, modified gravity theories adjust the strength of gravity on large scales and modify the weak gravitational lensing effect. Such alterations permit us to construct a model-independent observational test to ascertain the validity of the proposed theories \cite{measuringdarkside, intro46, intro47, intro48}. Model-independent methods, specifically at the linear perturbation level, were formulated in \cite{intro49,intro50, intro51, intro52, intro53}. These studies employ parameterization and principal component analysis to constrain the cosmic growth and expansion history. An alternative strategy involves utilizing the constraints and known characteristics of dark energy, along with its behavior, as a foundation for developing and evaluating theories. This approach has been elaborated upon in \cite{intro54, intro55, intro56}.

Each of the methods mentioned above requires the use of parameterizations to differentiate between various models based on data. However, it's essential to acknowledge that these parameterizations introduce biases that are contingent on the specific parameterization chosen. These biases have the potential to influence the accuracy of the results \cite{observabl-unobservable}.

On the other hand, considering the perturbations level, there is an approach that is fully model-independent. This measures properties like the anisotropic stress without making any assumptions about the nature of the dark energy. The existence of anisotropic stress results in the appearance of gravitational slip i.e. the slip parameter $\eta$ defined as the ratio between two gravitational potentials in a perturbed Friedmann-Robertson-Walker (FRW) universe being different from unity. Given that the contribution of neutrinos' free streaming to the anisotropic pressure in the late stages of the universe becomes negligible in the presence of perfect fluid matter, the only source of anisotropic stress could be a modification of gravity. In other words, the existence of anisotropic stress leads to a difference in values between the two scalar gravitational potentials that appear in the linear perturbation equations of motion and can be described as the gravitational slip parameter $\eta \equiv \frac{\Phi}{\Psi} \ne 1 $ \cite{observabl-unobservable, probing}. 

The advantage of the gravitational slip parameter is that it can be constructed based on model-independent observable quantities, which could potentially confirm or rule out a particular model\cite{model-independent-constraint, model-independent-eta}. Moreover, another advantage of this method is the relation between gravitational slip parameter and gravitational waves. The existence of anisotropic stress at the first order in perturbations caused by perfect-fluid matter leads to modifications in the propagation or behaviour of gravitational waves. In General Relativity, gravitational waves are the only propagating degrees of freedom. Therefore, as discussed earlier, it is logical to define modified gravity models as those in which gravitational waves are altered \cite{re7, re8, re9, re10}. Imperfect fluids with anisotropic stress may impact the two gravitational potentials but do not affect the propagation of tensor modes which are associated with gravitational waves \cite{anisotropic-stress, testingGW}. So far, the relation between gravitational slip parameter and non-standard gravitational waves in the context of scalar-tensor theories such as Horndeski theory has been studied \cite{nonstandardGW, HorndeskiGW}. 

Motivated by the above mentioned discussions, in this paper, we study the gravitational slip parameter and the behaviour of gravitational waves in the context of theories containing torsion, particularly the ECT as the simplest model of this class. In section \ref{sec:EC} we briefly review the definitions and field equations in ECT, and obtain the scale factor and the spin density in three different types of the universe including radiation-dominated era, matter-dominated era, and dark energy-dominated era. In section \ref{sec:eta} we investigate linear scalar cosmological perturbations in ECT in order to find gravitational slip parameter within this framework. In section \ref{sec:GW} we study the propagation of tensor modes (i.e., gravitational waves) in EC  cosmology and analyze the behavior of gravitational waves in this model. Section \ref{sec:conclusion} is devoted to our concluding remarks.

\section{\label{sec:EC}Einstein-Cartan cosmology\protect}

\subsection{\label{sec:fieldeq} Definitions and Field Equations}

Torsion tensor \footnote{In this paper we assume a flat FRW universe with the metric signature (+,-,-,-).} is defined as the anti-symmetric part of the connection 
\cite{bravo-2019}
\begin{equation}
    \label{1}
    S_{\mu \nu}\,^\alpha \,=\,\Gamma_{[\mu \nu]}^\alpha\,=\, \frac{1}{2}\,(\Gamma_{\mu \nu}^\alpha\,-\, \Gamma_{\nu \mu}^\alpha).
\end{equation}

In order to compare torsion theories with general relativity, it is suitable to divide the full connection in two parts
\begin{equation}
	\label{2}
	\Gamma_{\mu \nu}^\alpha\,=\, \overset {\circ}{\Gamma}\,^\alpha _{\mu \nu} \,-\, K_{\mu \nu}\,^\alpha,
\end{equation}
where
$\overset {\circ}{\Gamma}\,^\alpha _{\mu \nu}$ is standard Christoffel connection and $ K_{\mu \nu}\,^\alpha$ is known as the contorsion tensor which is related to Torsion tensor by the following relations
 \begin{equation}
 	\label{3}
 	K_{\mu \nu}\, ^\alpha = -S_{\mu \nu}\, ^\alpha + S_\nu \,^ \alpha \, _ \mu - S^\alpha \, _{\mu \nu},
 \end{equation}
\begin{equation}
	\label{4}
	S_{\mu \nu}\, ^\alpha = - K_{[\mu \nu]}\, ^\alpha = -\frac{1}{2}\,(K_{\mu \nu}\, ^\alpha - K_{\nu \mu}\, ^\alpha).
\end{equation}
For the remainder of this section we adopt the notation of reference \cite{bravo-2019}. Since the above mentioned connection \ref{2} is not necessarily symmetric, the Riemann tensor in ECT will have additional terms compared to GR  \footnote{Quantities with an over-circle indicate GR quantities and are obtained from the Christoffel connection}

\begin{equation}
	\label{6}
\begin{split}
	\mathrm{R}_{\mu \nu \alpha}\,^\beta &= \overset {\circ}{\mathrm{R}}_{\mu \nu \alpha}\,^\beta \,- \nabla_\mu K_{\nu \alpha}\, ^\beta \, - \nabla_\nu K_{\mu \alpha}\, ^\beta \,\\ & - K_{\nu \mu} \, ^\lambda K_{\lambda \alpha}\, ^ \beta \, + K_{\mu \nu} \, ^\lambda K_{\lambda \alpha}\, ^ \beta \, \\ & -K_{\nu \alpha}\, ^\lambda K_{\mu \lambda}\, ^\beta \, +K_{\mu \alpha}\, ^\lambda K_{\nu \lambda}\, ^\beta \\ & =\overset {\circ}{\mathrm{R}}_{\mu \nu \alpha}\,^\beta\, -2\nabla_{[\mu}K_{\nu] \alpha}\, ^\beta \,+2K_{[\mu \nu]}\,^\lambda K_{\lambda \alpha}\, ^\beta \,\\ & + 2K_{[\mu| \alpha}\,^\lambda K_{|\nu] \lambda}\, ^\beta,
\end{split}
\end{equation}
Contracting the Riemann tensor, the Ricci tensor and Ricci scalar will be 
\begin{equation}
	\label{7}
 \begin{split}
	\mathrm{R}_{\mu \nu} &= \overset {\circ}{\mathrm{R}}_{\mu \nu}\,-\nabla_\lambda K_{\mu \nu}\,^\lambda \,+\nabla_\mu K_{\lambda \nu}\,^\lambda \, \\ &+K_{\rho \mu}\,^\lambda K_{\lambda \nu}\,^\rho \,-K_{\mu \nu}\,^\lambda K_{\rho \lambda}\,^\rho,
\end{split}
\end{equation}

\begin{equation}
	\label{8}
\begin{split}
	\mathrm{R}&= \overset {\circ}{\mathrm{R}}\,+2 \nabla‌_\lambda K_{\alpha}\,^{\lambda \alpha} \,+ K_{\rho}\,^{\alpha \lambda} K_{\lambda \alpha}\,^{\rho}\,\\ &- K_{\alpha}\, ^{\alpha \lambda} K_{\rho \lambda}\,^{\rho}\\
	&=\overset {\circ}{\mathrm{R}}\,-4(\nabla_\lambda \,-S_{\alpha \lambda}\,^\alpha)\, S_\rho \,^{\lambda \rho} \,\\ &+2 S_{\alpha \beta \gamma} S^{\gamma \beta \alpha}\,+S_{\alpha \beta \gamma} S^{\alpha \beta \gamma}.
\end{split}
\end{equation}

The two field equations in Einstein-Cartan gravity are obtained by variation of the Lagrangian with respect to the metric and connection fields. One of the field equations gives an algebraic relation between the torsion and contorsion and their source, the spin tensor of the matter fields $\tau_{\alpha \nu\nu}$ as \cite{bravo-2019}
\begin{equation}
	\label{9}
	\begin{split}
		S_{\alpha \nu}\,^\mu = \kappa&(\tau_{\alpha \nu}\,^\mu \,+ \frac{1}{2}\delta_\alpha^\mu \tau _{\nu \lambda}\,^\lambda \,-\frac{1}{2}\delta_\nu ^\mu \tau_{\alpha \lambda}\,^\lambda),\\
		&-2S_{\alpha \lambda}\,^\lambda= \kappa \tau_{\alpha \lambda}\,^\lambda.
	\end{split}
\end{equation}
 
\begin{equation}
	\label{10}
	K_{\mu \nu}\,^\alpha =\kappa(-\tau_{\mu \nu}\,^\alpha \,+ \tau_\nu\,^\alpha\,_\mu \,-\tau^\alpha\,_{\mu \nu}\,-\delta_\mu ^\alpha \tau_{\nu \lambda}\,^\lambda \,+ g_{\mu \nu} \tau^\alpha\,_\lambda\,^\lambda).
\end{equation}

The other field equation of EC theory can be written as 
\begin{equation}
    \label{12}
\begin{split}
 	&\overset{\circ}{\mathrm{G}}_{\mu \nu}=\kappa \overset{\circ}{\mathrm{T}}_{\mu \nu} \,+\kappa^2 \bigg\{-4 \tau_{\mu \lambda}\,^{\alpha} \tau_{\nu \alpha}\,^{\lambda]}\,-2 \tau_{\mu \lambda \alpha} \tau_\nu\,^{\lambda \alpha}\,\\ & +\tau_{\alpha \lambda \mu} \tau^{\alpha \lambda}\,_\nu+\frac{1}{2}g_{\mu \nu} \big(4\tau_\lambda \,^\beta \,_{[\alpha} \tau^{\lambda \alpha}\,_{\beta]}\,+\tau^{\alpha \lambda \beta} \tau_{\alpha \lambda \beta}\big) \bigg\}
\end{split}
\end{equation}
where $\overset{\circ}{\mathrm{G}}_{\mu \nu}$ and $\overset{\circ}{\mathrm{T}}_{\mu \nu}$ refer to Einstein and Energy-Momentum tensors in GR respectively, and $\tau_{\mu \lambda}\,^{\alpha}$ indicates spin tensor. 

\subsection{\label{sec:a and q} Scale Factor and Spin Density}

We now use the two field equation above to determine background cosmological solutions for a single component universe. We begin by assuming the FRW metric for the universe
\begin{equation}
	\label{13}
	ds^2 = dt^2 \,- a^2 (t)\ \gamma_{ij} dx^i dx^j.
\end{equation}
In EC cosmology, the symmetries of FRW space-time lead to the non-zero component of spin and torsion tensors to be \cite{Goenner_1984}
\begin{equation}
	\label{14}
 \begin{split}
	& q(t)= \tau_{0 1 1} \ = \tau_{0 2 2} \ = \tau_{0 3 3} = -\tau_{A 0 A},\\
    & h(t)= S_{1 1 0} \ = S_{2 2 0} \ = S_{3 3 0} = -S_{A 0 A},
 \end{split}
\end{equation} 
where spin density function $q(t)$ and torsion function $h(t)$ are functions of cosmic time and are related to each other by the algebraic equation \ref{9}. Substituting \ref{13} and \ref{14}  into EC field equations, the Friedmann equations will be 
\begin{equation}
 	\label{15}
 	3\bigg(\frac{\dot{a}(t)}{a(t)}\bigg)^2 = \kappa \rho(t) \,+3 \kappa^2 a(t)^{-4}\ q(t)^2,
 \end{equation}
 
 \begin{equation}
 	\label{16}
  \begin{split}
 	& -\bigg(2\frac{\ddot{a}(t)}{a(t)} \,+(\frac{\dot{a}(t)}{a(t)})^2 \bigg) a^2(t) \gamma_{ij} = \\ & \bigg(\kappa a^2(t)\ p(t)\,- 2 \kappa^2 a^{-4}(t) q(t)^2 \bigg) \gamma_{ij}.
  \end{split}
 \end{equation}
where $a(t)$ and $q(t)$ are scale factor and spin density respectively, and $\kappa$ is Einstein gravitational constant $\kappa= 4 \pi $.
 Additionally, pressure and energy density of a perfect fluid are indicated  by $p(t)$ and $\rho(t)$, respectively. For a single component universe
\begin{equation}
	\label{17}
	\rho(t)= \frac{\rho_0}{a(t)^{3(1+\omega)}}.
\end{equation}
\begin{equation}
	\label{18}
	p(t)=\omega\rho(t)
\end{equation}
where $\omega$ is the equation of state parameter.
In following subsection we obtain the background scale factor and spin density as a function of time for various different matter content for the universe.
\subsubsection{\label{sec:a and q for RD} Radiation dominated universe \protect}
\begin{figure}[htbp]
\includegraphics[width=9cm, height=9cm]{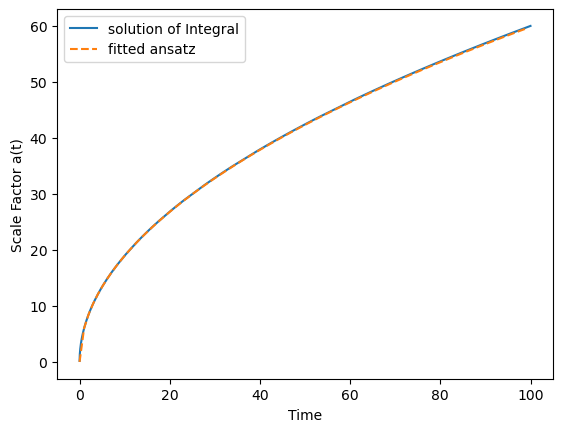}
\caption{\label{fig_1}The dashed line shows the ansatz (i.e.\ref{21}) and the solid line demonstrates the numerically obtained scale factor from the system of equations [\ref{15},\ref{16}]. The ansatz $a(t)=alpha t^{\beta}$ with fixed parameters as $\alpha=6,01$ and $\beta=\frac{1}{2}-0.001$ (i.e.\ref{21}) behaves exactly same to integral \ref{19}  which allows us to consider eq.\ref{21} as the scale factor of radiation dominated universe for the rest of our calculations.}
\end{figure}

By solving the system of equations \ref{15} and \ref{16} for $\omega=\frac{1}{3}$, we find the scale factor and spin density as follows
\begin{equation}
	\label{19}
	a(t)= Root \, of \Bigg(\int^{Z} \frac{3\ x}{\sqrt{x^3 \,+ \kappa \ \rho_0}} dx \, + t \Bigg),
\end{equation}
\begin{equation}
	\label{20}
	q(t)= \frac{ \sqrt{9\ \dot{a}(t)^2 \ a(t)^2\, - 3\ \kappa \rho_0}}{3\ \kappa}.
\end{equation}
The integral has no analytical solution and should be solved numerically. In order to obtain an analytical form for the background scale factor and spin density for future use, we solve the integral in \ref{19} numerically and try to fit an analytical function to the numerical solution. Since the effects of torsion on the cosmological dynamics in EC cosmology expected  to be small, we assume that the form the scale factor differ only slightly from its form in standard GR cosmology. We assume that for a radiation dominated universe, $a(t)\propto t^{\beta}$ where $\beta$ is a constant that differs only slightly from $1/2$. First, we solved the integral \ref{19} numerically by quadrature method using Python. Then we test the ansatz by directly substituting it into the system of equations \ref{15} and \ref{16}. At the end, the value of constant $\beta$ can be deduced by fitting an appropriate curve to the numerical solution. The result for $a(t)$ would be
\begin{equation}
	\label{21}
	a(t)\propto (t)^{\tfrac{1}{2}\, -0.001}.
\end{equation}
"Fig. \ref{fig_1}", shows the behavior of the fitted ansatz \ref{21} to numerical solution of the integral \ref{19}
 \footnote{Note that in this paper we take $\rho_0 = 9. 10^{-27}$
 and $\kappa = 2.077. 10^{-43}$}.

\begin{figure}[htbp]
\includegraphics[width=9cm, height=9cm]{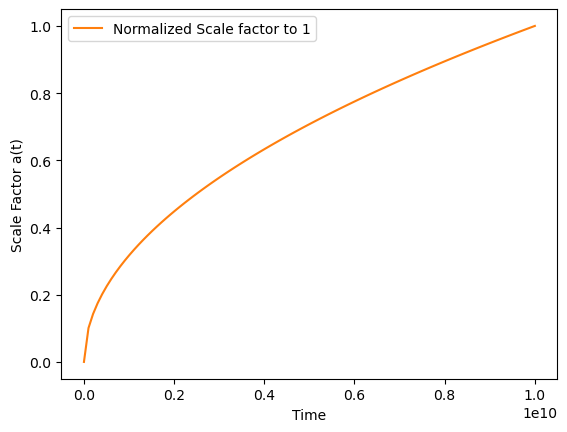}
\caption{\label{fig_2}Normalized scale factor for a radiation dominated universe in Einstein-Cartan cosmology. The present day scale factor $a_0$ was normalized to unity.}
\end{figure}

\begin{figure}[htbp]
\includegraphics[width=9cm, height=9cm]{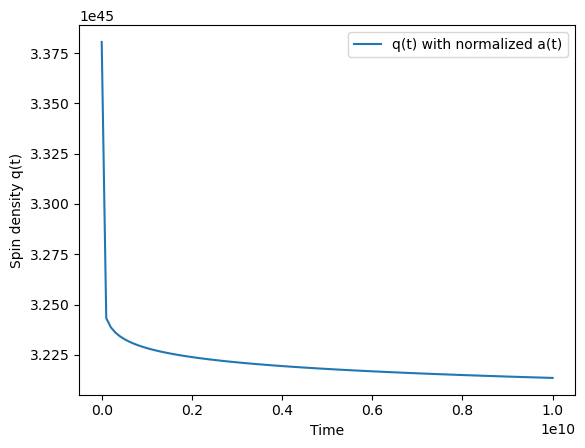}
\caption{\label{fig_3}  The graph indicates the spin density's value for the radiation dominated universe changes regarding time (eq.\ref{20}). The effects of torsion in very early universe is considerable.}
\end{figure}

Figure \ref{fig_2} shows the behaviour of $a(t)$ in (eq.\ref{20}) when the present day scale factor $a_0$ was normalized to unity. The behavior of the spin density in a radiation dominated universe, \ref{20} is indicated in "Fig. \ref{fig_3}".

\subsubsection{\label{sec:a and q for MD} Matter Dominated Universe \protect}
\begin{figure}[htbp]
\includegraphics[width=9cm, height=9cm]{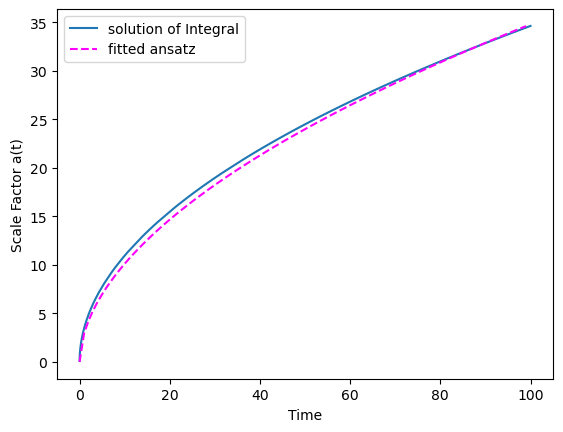}
\caption{\label{fig_4}The dashed line indicates the ansatz of scale factor (i.e.\ref{25}) and the solid line shows the obtained scale factor from the system of equations [\ref{15}-\ref{16}]. The ansatz (i.e.\ref{25}) behaves approximately the same as integral \ref{23} which allows us to consider eq.\ref{25} as the scale factor for matter dominated universe in the rest of our calculations.}
\end{figure}

\begin{figure}[htbp]
\includegraphics[width=9cm, height=9cm]{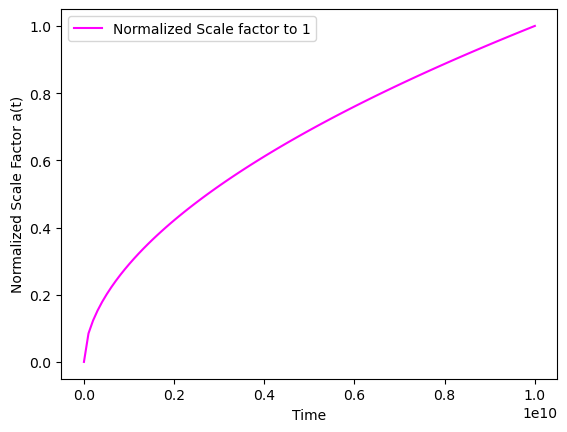}
\caption{\label{fig_5} Normalized scale factor in a matter dominated universe in EC cosmology. The present day scale factor ($t=1*10^{10}$) was set to unity.}
\end{figure}

\begin{figure}[htbp]
\includegraphics[width=9cm, height=9cm]{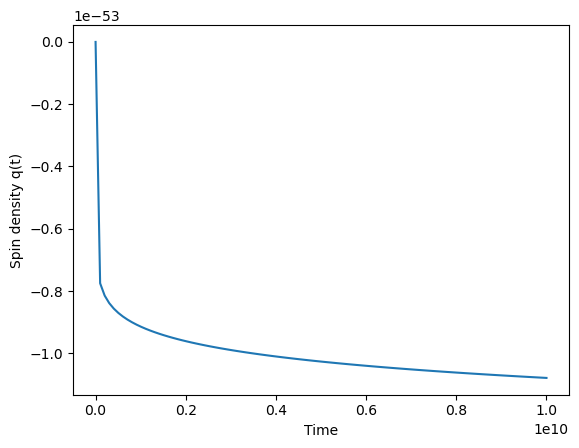}
\caption{\label{fig_6} Evolution of spin density for the matter dominated universe  (eq.\ref{24}) in EC cosmology. The torsion effects in this universe is infinitesimal.}
\end{figure}

By solving the system of equations \ref{15} and \ref{16} for $\omega=0$, the scale factor and spin density can be obtained for a matter dominated universe as

\begin{equation}
	\label{23}
	a(t)= Root \, of \Bigg(\int^{Z} \frac{x\ \sqrt{3}}{\sqrt{x(x^2 \,+ \kappa \ \rho_0)}} dx \, + t \Bigg),
\end{equation}
\begin{equation}
	\label{24}
	q(t)= \frac{- \sqrt{9\ \dot{a}(t)^2 \ a(t)^2\, - 3\ \kappa \rho_0 a(t)}}{3\ \kappa}.
\end{equation}
Using the same method as in the previous sub-section \ref{sec:a and q for RD}, we assume the scale factor to be in the form $a(t)\propto t^{\lambda}$ where the constant $\lambda$ differs only slightly from $2/3$. By fitting the ansatz to the numerical solution of integral \ref{23} we find 
\begin{equation}
	\label{25}
	a(t)\propto (t)^{\tfrac{2}{3}\, -0.13};
\end{equation}
"Fig. \ref{fig_4}", shows the behavior of the fitted ansatz \ref{25} to numerical solution of the integral \ref{23}.\\
The normalised scale factor of matter dominated universe is 
 depicted in "Fig. \ref{fig_5}".  "Fig. \ref{fig_6}" shows the evolution of the spin density function $q(t)$ in a matter dominated universe in EC cosmology.
The effect of torsion in a matter dominated universe is far less than in a radiation dominated universe.
 
\subsubsection{\label{sec:a and q for DD} Dark Energy Dominated \protect}
At late times universe is dominated by the cosmological constant\cite{Riotto}. By assuming $\omega = -1$ and solving the system of equations [\ref{15}-\ref{16}], we find the scale factor and spin density of a the dark energy dominated universe as
\begin{equation}
	\label{27}
	a(t)= \sqrt{2}\, +\, \sin\left(\frac{\sqrt{15}\ \sqrt{\kappa}\ \sqrt{\rho_0}}{3} \, t\right) \, +\,\cos \left( \frac{\sqrt{15}\ \sqrt{\kappa}\ \sqrt{\rho_0}}{3} \, t\right),
\end{equation}
\begin{equation}
	\label{28}
	q(t)= \frac{ a(t) \sqrt{ 9\ \dot{a}(t)^2 -3\ \kappa\rho_0 \ a(t)^2  } \, }{3 \ \kappa}.
\end{equation}

\begin{figure}[htbp]
\includegraphics[width=9cm, height=9cm]{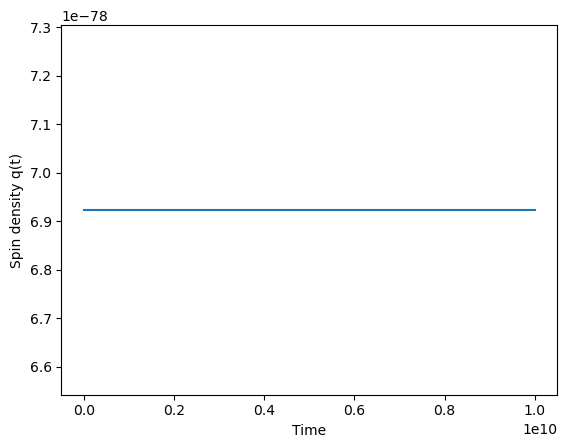}
\caption{\label{fig_8} The graph demonstrates the evolution of spin density for the dark energy dominated universe (eq.\ref{28}). The torsion effect in Einstein-Cartan theory for this universe is negligible.}
\end{figure}

\begin{figure}[htbp]
\includegraphics[width=9cm, height=9cm]{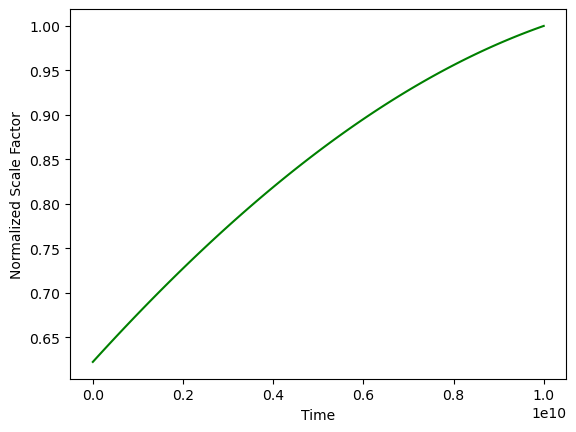}
\caption{\label{fig_9} Normalized scale factor for a Dark Energy dominated universe with torsion.}
\end{figure}

Behavior of the spin density (i.e. eq. \ref{28}) is plotted in "Fig. \ref{fig_8}" while "Fig. \ref{fig_9}" represents the behavior of the normalized scale factor in a dark energy dominated universe in EC cosmology.

\begin{figure}[htbp]
\includegraphics[width=9cm, height=9cm]{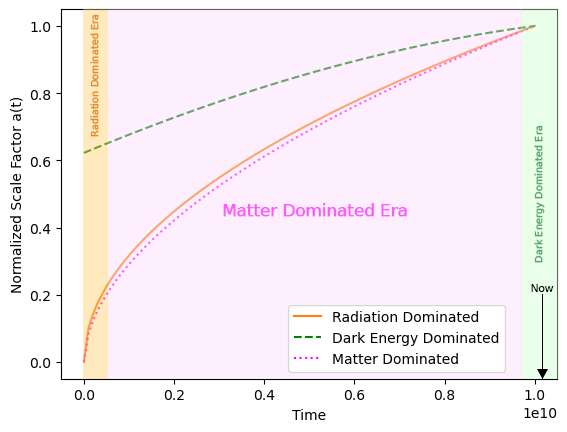}
\caption{\label{fig_10} The evolution of scale factor in  Einstein-Cartan cosmology for dark energy dominated universe \ref{27} (green dashed line), matter dominated universe\ref{25} (pink dotted line) and radiation dominated universe \ref{21} (orange solid line).}
\end{figure}

\begin{figure}[htbp]
\includegraphics[width=9cm, height=9cm]{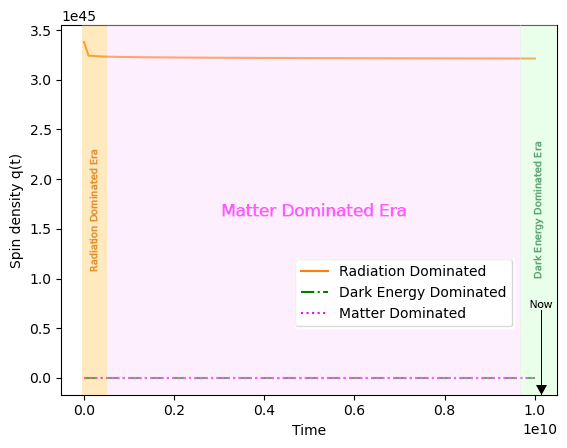}
\caption{\label{fig_11} Evolution of spin density in Einstein-Cartan cosmology for dark Energy \ref{28} (green dashed line), matter \ref{24} (pink dotted line) and radiation \ref{20} (orange solid line) dominated universes.}
\end{figure}

For a better comparison, we plotted the scale factor and spin density of the radiation, matter, and dark energy dominated universe in "Fig. \ref{fig_10}" and "Fig. \ref{fig_11}" respectively.

\section{\label{sec:eta} Slip Parameter \protect}
\subsection{\label{sec:eta-theory} Theoretical aspect of Slip Parameter}
The gravitational slip parameter $\eta$ is a direct measure of scalar (linear) anisotropic stress. At the linear level, the anisotropic stress $\eta$ defined as the ratio between the two gravitational potentials
\cite{probing, observabl-unobservable}
\begin{equation}
	\label{30}
	\eta(t,k) \equiv \Phi / \Psi.
\end{equation}

The existence of gravitational slip parameter ($\eta \ne 1$) in the presence of perfect fluid matter is a clue of a modification in gravity. In addition, this parameter is a model independent quantity which distinguishes the groups of gravity models \cite{probing, observabl-unobservable, brief-report}.

 We consider only the scalar degrees of freedom of the perturbed metric around a flat FRW background in the longitudinal (conformal Newtonian) gauge, where the line-element is of the form
 \begin{equation}
     \label{31}
     ds^2=(1+ 2'\Psi) dt^2- a(t)^2 (1-2 \Phi) \delta_{ij} dx^i dx^j,
 \end{equation}
where $\Psi \equiv \Psi(t,r)$ and  $\Phi \equiv \Phi(t,r)$ are the gravitational potentials and $a(t)$ is the scale factor. By plugging the perturbed metric \ref{31} and non-zero components of the spin tensor \ref{14} into the field equation of Einstein-Cartan theory \ref{12}; the time-time component \ref{32}, time-space component \ref{33} and space-space component \ref{34} of perturbed field equation of ECT can be written respectively as
\footnote{Here, and for the rest of calculations, an overdot denotes a time derivative, i.e. $\dot{a(t)}= \frac{d a(t)}{dt}$.}

\begin{equation}
	\label{32}
\begin{split}
	& 3\ H(t)^2 \,+ 2 \nabla^2 \Phi(t,r) \,- 6 \ H(t) \ \dot{\Phi} (t,r) \\ & \, = \kappa \Big[\rho(t) \,+ \delta \rho (t,r) + 2 \rho(t)\ \Psi(t,r) \Big] \\& \, + \kappa^2 \Big[3\ a(t)^{-4} \ q(t)^2\ + 12\ a(t)^{-4} \ q(t)^2 \ \Phi(t,r) \\& \, + 6 \ a(t)^{-4} \ q(t) \ Y(t,r) \ + 6\ a(t)^{-4} \ q(t)^2 \ \Psi(t,r) \Big],
\end{split}
\end{equation}

\begin{equation}
	\label{33}
	\begin{split}
	2 \partial_i \Big(\dot{\Phi} \ (t,r) \,+ H(t)\ \Psi(t,r) \Big) = 0,
	\end{split}
\end{equation}

\begin{equation}
	\label{34}
\begin{split}
	&\Bigg[ -2 \Big( \frac{\ddot{a}(t)}{a(t)} + \big( \frac{\dot{a}(t)}{a(t)}\big)^2 \Big) \ - \ \big( \frac{\dot{a}(t)}{a(t)} \big)^2 +  \nabla^2 (\Psi(t,r) \\ & - \Phi(t,r)) +\bigg( 4 \ \frac{\ddot{a}(t)}{a(t)} + 4 \big( \frac{\dot{a}(t)}{a(t)}\big)^2  + 2 \big( \frac{\dot{a}(t)}{a(t)}\big)^2 \bigg) \\ & \bigg( \Phi(t,r) + \Psi(t,r) \bigg) +  2 \ddot{\Phi}(t,r) \ + \ 2 H(t) \dot{\Psi}(t,r) \ \\ & + \ 4 H(t) \dot{\Phi}(t,r) \Bigg] \delta_{ij} +\ \partial_i \partial_j (\Phi(t,r)- \Psi(t,r)) \\ & =  \Bigg[ \kappa \ a(t)^2 P(t) \ +\ \kappa \ a(t)^2 \delta P(t,r) \\ & -\ 2 \kappa \ a(t)^2 P(t) \Phi(t,r)  - \ 25 \kappa^2 a(t)^{-2} q(t) Y(t,r) \\ & - 52 \kappa^2 a(t)^{-2} q(t)^2 \Phi(t,r) + 22 \kappa^2 a(t)^{-2} q(t)^2 \Psi(t,r) \\ & - \ 62 \kappa^2 a(t)^{-2} q(t)^2 \Bigg] \delta_{ij}.
\end{split}
\end{equation}

Note that we only consider scalar linear perturbation of the field equation (i.e. $\Pi_{ij} =0 $). Additionally, we only consider perturbation of the non-zero component of the spin tensor which is represented by $Y(t,r)$. $\delta \rho(t,r)$, $\delta P(t,r)$ and $H(t)$ refer to density perturbation, pressure perturbation and Hubble parameter, respectively \footnote{$\kappa$ is Einstein gravitational constant}
\begin{equation}
	\label{35}
	H(t)= \frac{\dot{a}(t)}{a(t)},
\end{equation}
\begin{equation}
	\label{36}
	\begin{split}
	&P(t)= \omega \rho(t),\\& \delta P(t,r)= \omega \delta \rho(t,r).
	\end{split}
\end{equation}

In the above equations, any scalar field $\chi$ can be expanded into Fourier modes
\begin{equation}
	\label{37}
	\chi = \int \frac{d^3 k}{(2 \pi)^3} \ e^{\mathit{i} \mathbf{k}.\mathbf{x}} \chi (t),
\end{equation}
performing the expansion, we can write the equation for the fluctuations as
\footnote{$\mathbf{k}$ is wave-number}
\begin{equation}
	\label{38}
\begin{split}
	& 3\ H(t)^2 \,- 2 \mathbf{k}^2 \Phi(t) \,- 6 \ H(t) \ \dot{\Phi} (t) \\ & = \kappa \Big[\rho(t) \,+ \delta \rho (t) + 2 \rho(t)\ \Psi(t) \Big] \\& + \kappa^2 \Big[3\ a(t)^{-4} \ q(t)^2  + 12\ a(t)^{-4} \ q(t)^2 \ \Phi(t) \\& + 6 \ a(t)^{-4} \ q(t) \ Y(t) + 6\ a(t)^{-4} \ q(t)^2 \ \Psi(t) \Big],
\end{split}
\end{equation}

\begin{equation}
 	\label{39}
 	\begin{split}
 		2 \mathbf{k} \Big(\dot{\Phi} (t) \,+ H(t)\ \Psi(t) \Big) =0,
 	\end{split}
\end{equation}

\begin{equation}
 	\label{40}
\begin{split}
 	&\Bigg[ -2 \Big( \frac{\ddot{a}(t)}{a(t)} + \big( \frac{\dot{a}(t)}{a(t)}\big)^2 \Big) \ - \ \big( \frac{\dot{a}(t)}{a(t)} \big)^2 \ - \ \mathbf{k}^2 (\Psi(t) + \Phi(t)) \\ & +\bigg( 4 \ \frac{\ddot{a}(t)}{a(t)} + 4 \big( \frac{\dot{a}(t)}{a(t)}\big)^2  + 2 \big( \frac{\dot{a}(t)}{a(t)}\big)^2 \bigg) \bigg( \Phi(t) + \Psi(t) \bigg) \\ &  +  2 \ddot{\Phi}(t) \ + \ 2 H(t) \dot{\Psi}(t) \ + \ 4 H(t) \dot{\Phi}(t) \Bigg] \delta_{ij} \\& +\ \mathbf{k}^2 (\Psi(t)- \Phi(t)) = \Bigg[ \kappa \ a(t)^2 P(t) \\ & +\ \kappa \ a(t)^2 \delta P(t) \ -\ 2 \kappa \ a(t)^2 P(t) \Phi(t) \\ & - \ 25 \kappa^2 a(t)^{-2} q(t) Y(t) \ -\ 52 \kappa^2 a(t)^{-2} q(t)^2 \Phi(t) \\ & +\ 22 \kappa^2 a(t)^{-2} q(t)^2 \Psi(t) \ - \ 62 \kappa^2 a(t)^{-2} q(t)^2 \Bigg] \delta_{ij}.
\end{split}
\end{equation}

In order to find the ratio of gravitational potentials, we shall insert the background scale factor and spin density obtained in the previous section  into perturbed field equations and solve the system of equations (\ref{38}, \ref{39}, \ref{40}).

\subsubsection{\label{sec:eta for RD} Radiation Dominated \protect}
In a radiation dominated universe, the scale factor can be written in terms of density as $a(t)= (\frac{\rho_0}{\rho(t)})^{1/4}$. As a result, spin density [\ref{20}] also can be written in terms of matter density
\begin{equation}
	\label{41}
	q(t)= \frac{ \sqrt{9\ (\frac{d}{dt} (\frac{\rho_0}{\rho(t)})^{1/4})^2 \ (\frac{\rho_0}{\rho(t)})^{1/2}\, - 3\ \kappa \rho_0}}{3\ \kappa}.
\end{equation}
Perturbation of previous equation \ref{41} gives us a relation between $Y(t)$ and $\delta \rho(t)$, where $Y(t) \propto (\frac{1}{\delta \rho(t)})^{1/14}$. Therefore, $Y(t)$ in perturbed field equations for radiation dominated universe is negligible.

\begin{figure}[htbp]
\includegraphics[width=9cm, height=9cm]{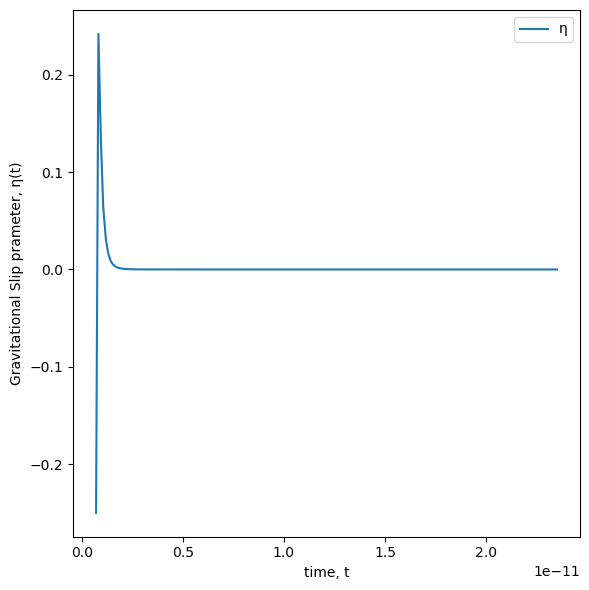}
\caption{\label{fig_12} The evolution of the  ratio of two gravitational potentials $\Phi(t)$ and $\Psi(t)$ (i.e. the Slip parameter) for a radiation dominated universe in ECT.}
\end{figure}

In sec [\ref{sec:a and q for RD}] we obtained background scale factor and spin density for radiation dominated universe in EC cosmology. In order to find $\Psi(t)$, $\Phi(t)$ and $\delta \rho(t)$ in this case, we substitute equations \ref{20} and \ref{21} into the system of equations [\ref{38}, \ref{39}, \ref{40}] and by assuming $\omega = 1/3$, solve the system of equations numerically.  The result for the evolution of gravitational slip parameter over time is illustrated  in "Fig. \ref{fig_12}".

\subsubsection{\label{sec:eta for MD} Matter Dominated \protect}
Same as previous sub-section \ref{sec:eta for MD}, using eq.\ref{18}, the  scale factor can be written in terms of density (i.e. $a(t)= (\frac{\rho_0}{\rho(t)})^{1/3}$). As a result, spin density [\ref{24}] can be written in terms of matter density as below
\begin{equation}
	\label{42}
	q(t)= \frac{- \sqrt{9\ (\frac{d}{dt} (\frac{\rho_0}{\rho(t)})^{1/3})^2 \ (\frac{\rho_0}{\rho(t)})^{2/3}\, - 3\ \kappa \rho_0 (\frac{\rho_0}{\rho(t)})^{1/3}}}{3\ \kappa}.
\end{equation}

By perturbing this equation, we found the relation between $Y(t)$ and $\delta \rho(t)$ as $Y(t) \propto (\frac{1}{\delta \rho(t)})^{2/22}- (\frac{1}{\delta \rho(t)})^{1/6}$. Therefore, $Y(t)$ in perturbed field equations for matter dominated universe will also be vanishingly small.

\begin{figure}[htbp]
\includegraphics[width=9cm, height=9cm]{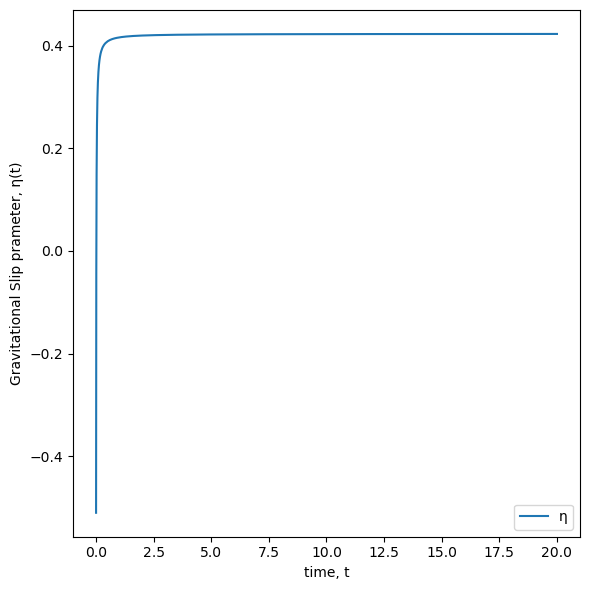}
\caption{\label{fig_13} The evolution of the  ratio of two gravitational potentials $\Phi(t)$ and $\Psi(t)$ (i.e. the Slip parameter) for a matter dominated universe in ECT.}
\end{figure}

In the next step, we insert the scale factor \ref{25}, and the spin density \ref{24} into the system of equations [\ref{38}-\ref{40}] and by assuming $\omega =0$, we Solve this system of equations numerically to find  $\Psi(t)$, $\Phi(t)$ and $\delta \rho(t)$. The result in the form of the ratio of gravitational potentials (i.e. slip parameter $\eta$) is demonstrated in "Fig. \ref{fig_13}". 

\subsubsection{\label{sec:eta for DD} Dark Energy Dominated \protect}
By the same analysis as the previous cases, we find that the perturbed spin tensor $Y(t)$ is also negligible in dark energy dominated universe.

\begin{figure}[htbp]
\includegraphics[width=9cm, height=9cm]{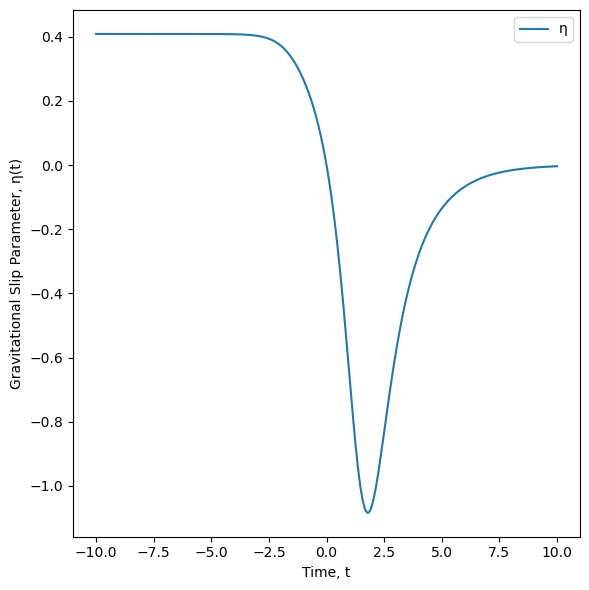}
\caption{\label{fig_14} The evolution of the slip parameter versus time for a dark energy dominated universe ECT.}
\end{figure}

Consequently, we substituted the scale factor \ref{27} and spin density \ref{28} of dark energy dominated universe into the system of equations [\ref{38}- \ref{40}] and solved it numerically to find the slip parameter. The behavior of slip parameter over time is demonstrated in "Fig. \ref{fig_14}".

\subsection{\label{sec:eta-observe} Slip Parameter from an observational viewpoint}
In observational approaches, it is generally necessary to adopt a model or make assumptions about the initial conditions of the Universe before attempting to estimate a quantity\cite{model-independent-eta}.
Several studies \cite{probing,observabl-unobservable,independent-model2,independent-model3,independent-model4} demonstrate the possibility of obtaining three model-independent quantities, denoted as $P1, P2, P3$, that effectively eliminate the influences of the primordial power spectrum's shape and galaxy bias, where $E=H/H_0$ and $z$ represents redshift. (see \cite{model-independent-eta} for more explanation).

In \cite{observabl-unobservable}, it has been shown that the ratio of the gravitational potentials and therefore the gravitational slip parameter can be reconstructed in a model-independent manner by pure observational quantities as following
\begin{equation}
	\label{43}
	\eta_{obs} \equiv \frac{3\ P_2 \ (1+z)^3}{2E^2 \ (P_3 +2 + \frac{E^{\prime}}{E})} -1 = \eta.
\end{equation}
$\eta_{obs}$ remains independent of any assumptions concerning the cosmic expansion. In other words, it does not rely on a specific $\Lambda$CDM background or any other particular model \cite{model-independent-eta}.

Using the above equation \ref{43}, the authors in reference \cite{model-independent-eta}, obtained the value of the slip parameter $\eta_{obs}$ in different Redshift using three different statistical methods. Their reported observational value of slip parameter is summarized in Table \ref{table_1}. Note that the values generally have great amount of uncertainties.

In order to compared our results for the slip parameter in Einstein-Cartan cosmology summarized in figures "Fig. \ref{fig_12}", "Fig. \ref{fig_13}" and "Fig. \ref{fig_14}", with the observational values presented in  Table \ref{table_1}, we used the standard relation between the redshift and the scale factor
\begin{equation}
    \label{44}
    1+z = \frac{a(t_0)}{a(t)},
\end{equation}
in order to convert cosmic times to redshifts and vice versa.  As can be seen from the values in the table, the results for Einstein-Cartan cosmology fall within the observational values obtained by using GaPP and Linear Regression methods.  

\begin{table*}[ht]
\caption{\label{table_1}The  first three rows show the observed value of slip parameter in different Redshifts using three different statistical methods \cite{model-independent-eta}. The last row is the  theoretically obtained value of slip parameter in each Redshift in the context of the Einstein-Cartan theory obtained in this paper. }
\begin{tabular}{|c|c|c|c|c|}
    \hline
    \hline
     \multicolumn{5}{c}{} \\
       \multicolumn{5}{c}{Observed value of Slip parameter} \\
    \hline
    Method & \quad \, \,  \, Choice of $H_0$ \quad \quad \quad & $\eta (z=0.294)$ & $\eta (z=0.58)$ & $\eta (z=0.86)$ \\
     \hline
     Binning & $H_0$ HST & $0.48 \pm 0.45 \,$ & $ −0.03 \pm 0.34 $ &  $\, −2.78 \pm 6.84$ \\
      & $H_0$ Planck & $ 0.56 \pm 0.54 \,$ & $ −0.14 \pm
     0.32$  &  $\, −6.75 \pm 75.64$ \\ \hline
     GaPP & $H_0$ HST & $\, 0.49 \pm 0.25$ & $0.94 \pm 0.33  $  &  $0.27 \pm 0.67\,$ \\
      & $H_0$ Planck & $\, 0.31 \pm 0.22 $ & $0.72 \pm 0.33$  &  $ 0.36 \pm 0.79\,$ \\ \hline
    Linear Regression & $H_0$ HST & $0.57 \pm 1.05\,$ &  $ 0.48 \pm 0.96 $  &  $ −0.11 \pm 3.21\,$ \\
      & $H_0$ Planck & $0.51 \pm 1.07\,$  & $0.37 \pm
      0.93$  &  $−0.18 \pm 3.11\,$\\
      \hline
      \hline
      \multicolumn{5}{c}{} \\
      \multicolumn{5}{c}{Theoretically obtained value of Slip parameter in ECT} \\
    \hline
    Einstein-Cartan cosmology  & $z$ & $ 0.294 $  & $ 0.58 $  &  $ 0.86$ \\
    & $\eta $ & $ 0.42 $ & $ 0.40 $  &  $ 0.02$ \\
    \hline
\end{tabular}
\end{table*}

\section{\label{sec:GW} Gravitational Waves}
In order to obtain the gravitational waves (GW), we only consider the tensor perturbation of the metric\cite{baumann_2022}
\begin{equation}
	\label{30.1}
	ds^2 = dt^2 \,- a^2 (t)\ (\gamma_{ij} + h_{ij}) dx^i dx^j,
\end{equation}
where $h_{ij}$ decompose to scalar, vector and tensor perturbation
\begin{equation}
	\label{31.1}
	h_{ij} = 2C \delta_{ij} + \partial_{<i} \partial_{j>} E +2 \partial_{(i} E_{j)} + 2E_{ij}.
\end{equation}
focusing on the last term  we have
\begin{equation}
	\label{32.1}
	ds^2 = dt^2 \,- a^2 (t)\ (\gamma_{ij} + 2 E_{ij}) dx^i dx^j.
\end{equation}
Using the above metric \ref{32.1} and finding the perturbed field equation of Einstein-Cartan theory \ref{12}, the equation relevant to the propagation of gravitational waves will be
\begin{equation}
    \label{33.1}
\begin{split}
    \Box E_{i j}(t,r)&=\kappa(a(t)^2 P(t) + a(t)^2 \delta P(t,r))\delta_{i j}\\
     &+\kappa a(t)^2 \Pi_{i j}
     + \kappa^2 18 a(t)^{-2} q(t) Y(t,r) \gamma_{ij} \\
     &+ \kappa^2 2 a(t)^{-2} q(t)^2 E_{i j} (t,r),
\end{split}  
\end{equation}
where the d’Alembertian operator is
\begin{equation}
    \label{34.1}
    \Box E_{i j}= \partial_i \partial^j E_{i j}.
\end{equation}
We aim to obtain gravitational waves in vacuum, consequently,  the pressure and its perturbation ($P(t)$ and $\delta P(t,r)$) vanish. Also, anistropic stress ($\Pi_{i j}$) and perturbation of spin tensor ($Y(t,r)$) are negligible in cosmic scales. These considerations give us the equation for the Plane gravitational waves in EC theory as 
\begin{equation}
    \label{35.1}
    \Box E_{i j}(t,r)= 2 \kappa^2 a(t)^{-2} q(t)^2 E_{i j} (t,r).
\end{equation}
GW for radiation dominated and matter dominated universe in the EC theory is studied in detail in following sub-sections.
 
\subsubsection{\label{sec:gw for RD} Radiation dominated universe\protect}
By inserting scale factor \ref{21} and spin density \ref{20} into eq.\ref{35.1} and solving the equation, the GWs in radiation dominated universe can be obtained as 
\begin{equation}
\begin{split}
\label{37.1}
    & E(t)= e^{-i m t} t \bigg( C_1 \,  _1F_1[
    1-\frac{l}{2 i m},2, 2imt]\\
    & + C_2 \, U[1-\frac{l}{2 i m},2, 2imt] \bigg),
\end{split}
\end{equation}
where $l= 17$, and $m$ is wave-number due to the Fourier transformation \ref{37}. Additionally,  $_1F_1(a,b,x)$ and $U(a,b,x)$ are Confluent Hypergeometric Functions of first  and second kinds, respectively \cite{Confluent}.
The confluent hypergeometric function can be expanded in hypergeometric series \cite{arfken, wittaker}.

\begin{figure}[ht]
\includegraphics[width=8cm, height=6cm]{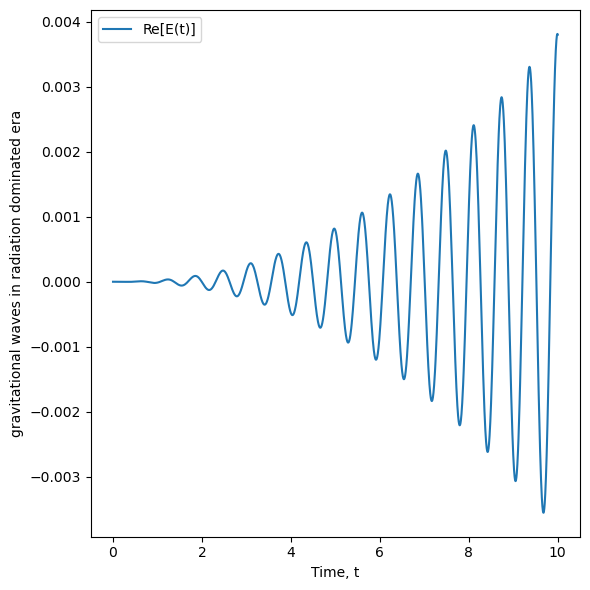}
\caption{ The graph shows the behavior of the normalized gravitational waves regarding time in the context of EC theory for radiation dominated era. Note that we consider m=10.}
\label{fig_15}
\end{figure}

The behavior of gravitational waves for radiation dominated era in the EC theory is illustrated in Fig[\ref{fig_15}].\\ 

\subsubsection{\label{sec:gw for MD} Matter dominated universe\protect}
\begin{figure}[ht]
\includegraphics[width=8cm, height=6cm]{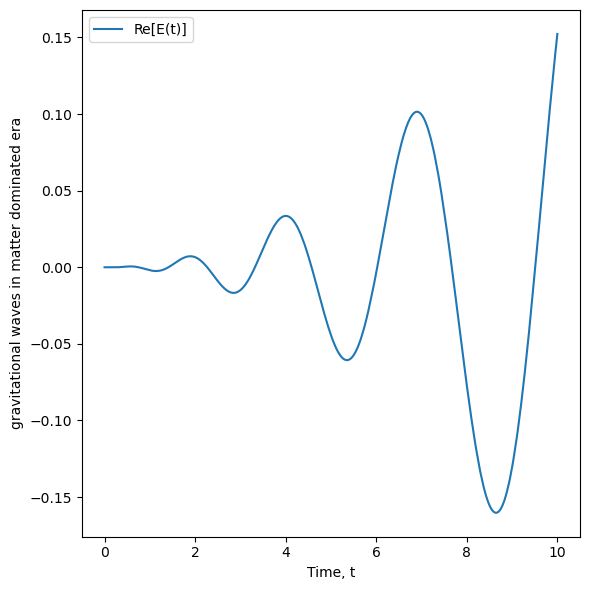}
\caption{\label{fig_16}The graph shows the behavior of gravitational waves regarding time through the EC theory for the matter dominated era. Note that we consider m=10.}
\end{figure}

By substituting scale factor \ref{25} and spin density \ref{24} of matter dominated universe into the GWs equation \ref{35.1}, the GWs in matter dominated universe will be
 \begin{equation}
     \label{39.1}
     \begin{split}
     E(t) &= C_{1} \, M[-\frac{i l_2}{2 m},-\frac{1}{2} i \sqrt{4 l_1-1}, 2 i m t] \\
     &+C_{2} \, W[-\frac{i l_2}{2 m},-\frac{1}{2} i \sqrt{4 l_1-1}, 2 i m t],
     \end{split}
 \end{equation}
where $l_1=4.8$ and $l_2= 2.9$. The Whittaker functions (i.e. $M$ and $W$) can be written in terms of confluent hypergeometric functions of  first  and second kinds \cite{Abramowitz-Stegun, Whittaker-Watson} as below
\begin{equation}
     \label{40.1}
     \begin{split}
         &E(t)= e^{-i m t} (2 i m t)^{\frac{1}{2} \left(1-i \sqrt{4 l_1-1}\right)}\\
         &\Bigg[ C_{1}  \,  _1F_1\left(\frac{i l_2}{2 m}+\frac{1}{2} \left(1-i \sqrt{4 l_1-1}\right),1-i \sqrt{4 l_1-1},2 i m t\right)\\
         &+C_{2}\, U \left( \frac{i l_2}{2 m}+\frac{1}{2} \left(1-i \sqrt{4 l_1-1}\right),1-i \sqrt{4 l_1-1},2 i m t \right) \Bigg].
     \end{split}
 \end{equation}
Using power series expansion of hypergeometric functions,
behavior of gravitational waves for the matter dominated universe is depicted in Fig[\ref{fig_16}].

\section{\label{sec:conclusion} Conclusion \protect}
 The presence of anisotropic stress introduces a disparity in the values of two scalar gravitational potentials, $\phi$ and $\psi$, which appear in the first-order perturbation equation of motion. The gravitational slip parameter, measuring the anisotropic stress, is defined as the ratio of these two scalar gravitational potentials $\eta =\frac{\phi}{\psi}$. Given that the existence of anisotropic stress and gravitational slip parameter ($\eta \ne 1$) in the presence of perfect fluid matter could signify a modification of gravity, the gravitational slip parameter serves as a potential indicator of such modifications.

Furthermore, the gravitational slip parameter can be constructed based on model-independent observable quantities, providing a fully model-independent approach to testing modified gravity theories. In this paper, we applied this methodology to test Einstein-Cartan theory (ECT), a simple theory incorporating torsion.

By deriving the equations of motion within the framework of ECT and considering the Friedmann-Robertson-Walker metric for the universe, we obtained the scale factor and spin density for three distinct single-component universes: radiation-dominated, matter-dominated, and dark energy-dominated. Studying the linear scalar perturbations with a  non-zero spin tensor in the ECT, we find the behavior of the slip parameter over time in the context of ECT. The comparison of the theoretical values of the slip parameter within the ECT framework with the observed values provides evidence supporting the possibility of modified gravity in the sense of ECT.

Another test for the modification of gravity is the generation and propagation of cosmic gravitational waves. As modified gravity theories alter the behavior of gravitational waves, we can infer a connection between the modifications of anisotropy constraints and gravitational wave propagation. Our analysis demonstrates that introducing spin density to the matter part of the field equations and modifying the source term influence the behavior and amplitude of gravitational waves.

\newpage
\providecommand{\noopsort}[1]{}\providecommand{\singleletter}[1]{#1}%

\end{document}